\documentclass[aps,pra,twocolumn,superscriptaddress]{revtex4-1}
\usepackage{amsfonts}
\usepackage{amsmath}
\usepackage{amssymb}
\usepackage{graphicx}
\usepackage{times}
\usepackage{physics}
\usepackage[colorlinks=true,linkcolor=blue,citecolor=red,plainpages=false,pdfpagelabels]
{hyperref}
\usepackage{color}

\providecommand{\U}[1]{\protect\rule{.1in}{.1in}}

\begin{document}

\title{Discrete-Time Modelling of Quantum Evolutions, the Energy-Time Uncertainty Relation and General Extensions in the Entangled History Formalism}
\author{Hai Wang}
 \email{3090101669@zju.edu.cn}
 \affiliation{School of Mathematical Sciences, Zhejiang University, Hangzhou 310027, PR~China}
 \author{Ray-Kuang Lee}
 \email{rklee@ee.nthu.edu.tw}
\affiliation{Institute of Photonics Technologies, National Tsing Hua University, Hsinchu 30013, Taiwan}
\affiliation{Physics Division, National Center for Theoretical Sciences, Hsinchu 30013, Taiwan}
 \author{Junde Wu}
 \email{Corresponding author: Junde Wu (wjd@zju.edu.cn)}
 \affiliation{School of Mathematical Sciences, Zhejiang University, Hangzhou 310027, PR~China}

\date{\today}
\begin{abstract}
Time evolution is an indivisible part in any physics theory. Usually, people are accustomed to think that the universe  is a fixed  background and the system itself evolves step by step in time. However, Yakir Aharonov challenges this view using his two-vector formalism. In this paper, using the entangled history formalism, we attain three achievements. Firstyl, we give an affirmative  answer to Yakir Aharonov's question. Secondly, we reveals the energy-time uncertainty relationship from two extreme cases. Thirdly, we generalize previous  methods to quantum channel and density matrix, not just unitary evolutions or pure states.
\end{abstract}

\maketitle

\section{Introduction}
Since physics is born, statics and dynamics are two cores of physical theory. One is focused on the description of a system at a fixed instant, the other is concerned with the time evolution of a system under a theory framework. In quantum mechanics, we can calculate expectations of observables of a system at any single instant to know the system. And using the Schrodinger equation, we can know how the system evolves in time. Eigenvalues and expectations of hermitian operators play a very important role in the quantum mechanics theory. In fact, this viewpoint shows no difference with the canonical Newton form, where we can get information of systems at any fixed instant and know its evolution by the Newton formula. However, there is a subtle problem. Imagine, in classic physics, there is a particle, whose trajectory is $x(t)$ during a period $(0,\ T]$ predicted by theory. Now if we choose $N$ different instants $t_{1},\ t_{2},...\ t_{N}$ from $(0,\ T]$. Then from classic physics, it is  obvious that we can prepare $N$ particles of the same kind and then at some given time $\tau$ measure these particles, we will get the information as same as got by measuring the original particle at $t_{1},\ t_{2},...\ t_{N}$. Then a natural  question arises, wether we can evolutions in quantum mechanics in the same pattern. That is, for a particle in quantum mechanics, assume that it evolves in the time  $(0,\ T]$, then whether we can find something in $H=H_{N}\otimes \ ...\otimes H_{2}\otimes H_{1}$ to represent the information of the original particle at instants $t_{1},\ t_{2},...\ t_{N}$ from $(0,\ T]$. That is what Aharonov asks in his paper \cite{yak}.

\section{Differences and Solutions of Aharonov }

\subsection{Differences between classic physics and quantum mechanics}

 Consider that there is a state $\ket{\phi}\in H_{0}$ at the instant $t_{0}$, and in the following period  $(0,\ T]$, it goes through the trivial evolution. Then in this case ,during this period $(0,\ T]$ , the evolution is $$U= I.$$ Above all, if we choose $N$ different instants $t_{1},\ t_{2},...\ t_{N}$ from this period, then we may use $$\ket{\phi}\otimes \ ... \otimes \ket{\phi}\in \otimes_{i=0}^{N}H_{i},$$ to represent the case of the original particle at instants  $t_{1},\ t_{2},...\ t_{N}$, where $$H_{i}=H_{0}, \forall i\in \{0,...\ N\}.$$  But that is wrong. Contradictions are expressed mainly in two aspect. One is that, this modelling tells too much information. And the other is that this modelling tells too little. Suppose that we are given an unknown state $\ket{\phi}$, then we have no way to decide which state this state actually is. However, if we are given $\otimes_{i=0}^{N} \ket{\phi_{i}},$ where $$\ket{\phi_{i}}=\ket{\phi}, \forall 0\leq i\leq N.$$ Then although the state $\ket{\phi}$ is unknown, "by making different measurements on the different copies and looking at the statistics of the results we can learn the state" \cite{yak}. And this approximation will be better and better as $N$ becomes larger. That is the first problem.  On the other hand, if there is a state $\ket{\phi}\in C^{2}$ and it goes through the trivial evolution during the time $(0,\ T]$. Then using multi-time variables like $$\sigma_{\overrightarrow{r}}(t_{2})- \sigma_{\overrightarrow{r}}(t_{1}),$$ where $\sigma_{\overrightarrow{r}}$ is some spin operator and $$t_{1},\ t_{2}\in (0,\ T],$$ we can get strange results. That is, no matter which direction we choose, we will always get $$\sigma_{\overrightarrow{r}}(t_{2})- \sigma_{\overrightarrow{r}}(t_{1})=0.$$ Concretely, in the von Neuman's measuring formalism, if we make $q$ the pointer position of the measuring device and $p$ the conjugate momentum of the measuring device \cite{yak}. Then the above measuring process can be described by the interaction Hamiltonian $$H_{int}= -\delta(t-t_{1})p\otimes\sigma_{\overrightarrow{r}}+\delta(t-t_{2})p\otimes\sigma_{\overrightarrow{r}}.$$ Following reasoning in \cite{yak}, we will get $$\frac{dq}{dt}=i[q, H_{int}]=(\delta(t-t_{2})-\delta(t-t_{1}))\sigma_{\overrightarrow{r}}(t).$$ Then by a simple calculation, we will get $$q(t_{2}+\epsilon)-q(t_{1}+\epsilon)=\sigma_{\overrightarrow{r}}(t_{2})-\sigma_{\overrightarrow{r}}(t_{1}).$$
 So the difference between the final and initial positions of the pointer is completely dependent on the value of the two-time  observable $\sigma_{\overrightarrow{r}}(t_{2})- \sigma_{\overrightarrow{r}}(t_{1})$. The result of this experiment only tells the value of $\sigma_{\overrightarrow{r}}(t_{2})- \sigma_{\overrightarrow{r}}(t_{1})$ but not the value of $\sigma_{\overrightarrow{r}}(t_{1})$ or $\sigma_{\overrightarrow{r}}(t_{1})$ separately. Because the Hamiltonian acting on the spin is  zero, then we  will have $$\sigma_{\overrightarrow{r}}(t_{1})=\sigma_{\overrightarrow{r}}(t_{2}),$$ which results in $$q(t_{2}+\epsilon)-q(t_{1}+\epsilon)=\sigma_{\overrightarrow{r}}(t_{2})-\sigma_{\overrightarrow{r}}(t_{1})=0.$$ Details of the above reasoning is in \cite{yak}.

 Roughly speaking, in the above setting, if we first measure the observable  $\sigma_{\overrightarrow{r}}$ at the instant $t_{1}$, and then measure the same observable at the instant $t_{2}$, in the end, these two measurements must give the result, no matter what the result is. Although we cannot predict which result is detected if the state $\ket{\phi}$ is not an eigenvector of $\sigma_{\overrightarrow{r}}$, this fact is predicted by quantum mechanics. However, it is difficult to imagine that we can always get the same result for measuring some spin $\sigma_{\overrightarrow{r}}$ twice on different subsystems of $$\ket{\phi}\otimes \ ... \otimes \ket{\phi}\in \otimes_{i=0}^{N}H_{i},$$ where $$H_{i}=C^{2}, \forall i\in \{0,...\ N\}.$$ This shows that the above modelling cannot represent the temporal correlation hidden in this setting. 

\
 
 So in conclusion, there are two problems in the $\ket{\phi}\otimes \ ... \otimes \ket{\phi}$ model:
 
 \
 
 1. It tells too much. From this model, we can use quantum tomography to know what the unknown state it is. However, this  is  not allowed in the original  experiment setting.
 
 \
 
 2. It tells too little. From this  model, we cannot get the internal correlation between different instants.
 
\

 Based on these two considerations, \cite{yak} concludes that it is  impossible to find a simple vector in $\otimes_{i=0}^{N}H_{i}$ to represent a state's evolution at $N$ different instants.  However, later we will show this strategy actually works. Using the entangled history method, in fact, we can do this.

\subsection{Solution of Aharonov}
Again, assume that we have a qubit in a state $\ket{\phi}\in H_{0}=C^{2}$ and it evolves under the trivial way for time $(0,\ T]$. Now  we choose  $N$ instants $t_{1},...\ t_{N}$ from $(0,\ T]$. The discrete model,$$\ket{\phi}\otimes \ ... \otimes \ket{\phi}\in \otimes_{i=0}^{N}H_{i},$$ fails from the above discussions. The main problem is that, from this evolution setting, states of different instants are not independent from each other. For example, the state of instant $t_{i+1}$ is evolved from the state of instant $t_{i}$ trivially, where $$0\leq i\leq N-1.$$ So, how to express the evolution is the key problem. In their work \cite{yak}, using the two-vector formalism \cite{yak2}, they show that if we want to choose $N$ instants from the evolution time and try to use  something about the $N$ instants to represent the original particle's evolution, then for the above case, it must be of the following form$$\Phi_{N,N-1}^{\tau_{-},\tau_{+}}...\Phi_{2,1}^{\tau_{-},\tau_{+}}\Phi_{1,0}^{\tau_{-},\tau_{+}}\ket{\phi},$$ where $$\Phi_{k+1,k}^{\tau_{-},\tau_{+}}=\sum_{i}\ket{i}_{k+1\ k}^{\tau_{-}\ \tau_{+}}\bra{i}, \ \forall \ 0\leq k\leq N-1.$$ They call the construction $\Phi_{k+1,k}^{\tau_{-},\tau_{+}}$ the maximally entangled two-time state, which uses the two-vector formalism langugage. 

Just take a two-instant case as an example. Suppose we have a state $\ket{\phi_{0}}\in H_{0}$ at the initial time $t_{0}$. Then the system goes through the trivial evolution $I$ from $t_{0}$ to $t_{1}$. Usually, we  will show that the state at $t_{1}$ is $$\ket{\phi_{1}}=I \ket{\phi_{0}}=\ket{\phi_{0}}\in H_{1},$$where $H_{1}=H_{0}$. However, using the two-vector formalism, \cite{yak} sees discrete instants as time bricks. That is, each instant has two ends, one towards the past, one towards the future. Mathematically, for a fixed instant $H_{i}$, ket forms of vectors represent information towards the future and bra forms of vectors represent absorption of information from the past. Now back to our simple model. Suppose we have orthonormal basis $\{\ket{i}\}$ for $H_{1}=H_{0}$. Then the evolution between these two instants $t_{0}$ and $t_{1}$ can be expressed as $$I=\sum_{i}\ket{i}\bra{i}.$$ In \cite{yak}, the evolution will be expressed as $$\sum_{i}\ket{i}_{1,\ 0}^{\tau_{-},\ \tau_{+}}\bra{i},$$ where $\{^{\tau_{+}}_{0}\bra{i}\}$ functions  as  absorbing information from the instant $t\leq t_{0}$ and $\{\ket{i}_{1}^{\tau_{-}}\}$ functions as sending information from the instant $t_{1}$ to the future. Then from this explanation, $$I \ket{\phi_{0}}=\sum_{i}\ket{i}\bra{i} \ket{\phi_{0}}=\sum_{i}\ket{i}_{1,\ 0}^{\tau_{-},\ \tau_{+}}\bra{i}\ket{\phi_{0}}=\Phi_{1,0}^{\tau_{-},\tau_{+}}\ket{\phi_{0}}$$ actually  tells us that our initial state is $\ket{\phi_{0}}$ at $t_{0}$ and it goes through the trivial evolution to come to the instant $t_{1}$.

So actually, the maximally entangled two-time state $\Phi_{k+1,k}^{\tau_{-},\tau_{+}}$ is just another name of the unitary operator $I$ between instants $t_{k}$ and $t_{k+1}$. In \cite{yak}, they use  this construction to represent the temporal correlation hidden in the state's evolution and successfully solve the two problems in the model $$\ket{\phi}\otimes \ ... \otimes \ket{\phi}\in \otimes_{i=0}^{N}H_{i}.$$

To see more about the two-vector formalism, references \cite{yak2, yak3} are quite helpful.

\

\section{Our Results}

\subsection{Introduction to the Entangled History}

The entangled  history formalism, created by Jordan Cotler and Frank Wilczek \cite{frank1}, gives another viewpoint to see states' evolution in quantum theory. As the discrete form of the famous Feymann's path integral,  the core of this theory framework is that we can use the tensor product structure of Hilbert spaces to represent the evolution process of a system at different instants $t_{0}, t_{1},..., \ t_{N}$. Recently, using this formalism, they restate the Leggett-Garg inequality \cite{LG}, and show that the temporal correlation in quantum theory is the result of the superposition of states' evolution paths, which can be verified experimentally \cite{frank2, frank3}.

\

So from the above, it seems that analysis in \cite{yak} challenges the kernel of the entangled history formalism. Now using techniques in \cite{frank4}, we will show that the entangled history formalism still works and can give better description of discrete modelling of evolutions.

\

In the entangled history  formalism, if a system in $H_{0}$ evolves through the evolution $U$ in the time interval $[0, T]$, suppose we picks $n$ instants from the interval to see this evolution path, then this evolution path is seen as an element of the Hilbert space$$H=\odot_{i=0}^{n}H_{i},$$where every $H_{i}$ represents  the system at the instant $t_{i}$.

Take two-instant as an example. If we have a system in a state $\ket{\phi}$ at the instant $t_{0}$ and at the instant $t_{1}$ the system is in the state $\ket{\psi}$, furthermore we assume that the evolution during these two instants is the unitary operator $U$, then in the entangled history method, this will be described as $$\ket{\psi} \odot \ket{\phi},$$ with the bridge operator being $U$. Among these, the signature $\odot$ is as same as $\otimes$ mathematically, emphasizing its temporal nature.  In \cite{yak}, to encode the evolution information into the state's description, they use the so-called the maximally entangled two-time state, $$\Phi_{k+1,k}^{\tau_{-},\tau_{+}}==\sum_{i}\ket{i}_{k+1\ k}^{\tau_{-}\ \tau_{+}}\bra{i},$$ to connect different instants. However, if we choose an orthonormal base $\{\ket{m}\}$ for the Hilbert space $H_{0}$ and an orthonormal base for the Hilbert space $H_{1}$, then the unitory operator between the two instants $t_{0}$ and $t_{1}$ can be described by a matrix $(u_{nm})_{n,m}$, where $$u_{nm}=\bra{n}U\ket{m}.$$

Suppose that the initial state of the system at the instant $t_{0}$ is $$\ket{\phi}=\sum_{m}\alpha_{m}\ket{m},$$then  it goes under the evolution $U$ to the instant $t_{1}$. By \cite{frank4}, this can be described as $$\sum_{m,n}u_{nm}\alpha_{m}\ket{n}\odot \ket{m}.$$ Usually this will cause the entanglement in time and using the monitor systems method in \cite{frank4}, we can set up corresponding experiments to detect this kind of entanglement.

Following example is studied in detail in  \cite{frank4}. Again, let us back to the qubit case. Suppose at instant $t_{1}$ we choose a preferred orthonormal basis $\textit{B}_{1}= \{\ket{a},\ket{a_{\perp}}\}$ and at instant $t_{2}$  we choose $\textit{B}_{2}= \{\ket{b},\ket{b_{\perp}}\}$. If the initial state of our system is $\ket{s_{1}}=\alpha\ket{a}+\beta\ket{a_{\perp}}$. Then the history of our system for these two instants $t_{1}$ and $t_{2}$ can be expressed as

$$\ket{\Psi}=\textit{A}(a\rightarrow b)\alpha\ket{b}\odot \ket{a}+\textit{A}(a\rightarrow b_{\perp})\alpha\ket{b_{\perp}}\odot \ket{a}$$
$$+\textit{A}(a_{\perp}\rightarrow b)\beta\ket{b}\odot \ket{a_{\perp}}+\textit{A}(a_{\perp}\rightarrow b_{\perp})\beta\ket{b_{\perp}}\odot \ket{a_{\perp}},$$

 where $\textit{A}(a\rightarrow b)=\bra{b}U\ket{a}$ is the amplitude to transition
from $a$ to $b$ (and similarly for the other terms).

Now to access this history in experiment, \cite{frank4} introduces monitor systems. In  the above example, for the qubit system in the two-instant setting, before the evolution, we can couple our system to a two-qubit system initialized in the state $\ket{00}$. At time $t_{1}$, we apply a controlled unitary gate which makes the first monitor qubit $\ket{a}$ if the state of the main spin is $\ket{a}$, and $\ket{a_{\perp}}$ if it is $\ket{a_{\perp}}$. At time $t_{2}$, after the unitary time evolution $U$ has been applied to the main system, we apply a controlled unitary gate which makes the second monitor qubit $\ket{b}$ if the state of the main spin is $\ket{b}$, and $\ket{b_{\perp}}$ if it is $\ket{b_{\perp}}$. Then finally the state of the whole system will be $$\textit{A}(a\rightarrow b)\alpha\ket{b}_{main}\ket{b}\otimes \ket{a}+\textit{A}(a\rightarrow b_{\perp})\ket{b_{\perp}}_{main}\ket{b_{\perp}}\otimes \ket{a}+$$$$\textit{A}(a_{\perp}\rightarrow b)\beta\ket{b}_{main}\ket{b}\otimes \ket{a_{\perp}}+\textit{A}(a_{\perp}\rightarrow b_{\perp})\beta\ket{b_{\perp}}_{main}\ket{b_{\perp}}\otimes \ket{a_{\perp}}.$$  Projecting the main system onto $\frac{1}{\sqrt{2}}(\ket{b}+\ket{b_{\perp}})$ and then tracing it out, the monitor qubits exactly equal the history state $\ket{\Psi}$, but with $\otimes$'s instead of $\odot$'s. The above  procedure can be easily generalised to complex cases. In conclusion, by monitor systems, we can use them to track systems' evolution and uses the spatial correlation among minitor systems to store the temporal correlation arisen in the system's evolution. Note that in transformations from the main system's evolution temporal structure into monitor systems' spatial structure, the correspondence is the core. For the above example, although we can measure the monitor systems in any way we like, only when the eigenvectors of the measurement are the superpositions of $$\{\ket{b}, \ket{b_{\perp}}\} \otimes \{\ket{a}, \ket{a_{\perp}}\},$$ we can make inferences about the main system's evolution path from measurements' results. That is why this formalism has somewhat feelings of coherence.

\subsection{Answer to problems of Aharonov}

Now, back to our initial problem. Suppose we have a qubit in a state $\ket{\phi}\in C^{2}$ and it goes through the trivial evolution $I$ for some time. Let $$t_{1}, t_{2},...\ t_{N}$$ be our chosen instants. If we choose eigenvectors of some spin $\sigma_{\overrightarrow{r}}$ $$\{\ket{0}_{\overrightarrow{r}}, \ket{1}_{\overrightarrow{r}}\}$$as the orthonormal base for Hilbert spaces $$H_{i}=C^{2},\ 0\leq i\leq N,$$ then the initial state of the system can be expressed as $$\ket{\phi}=\alpha_{0}\ket{0}_{\overrightarrow{r}}+\alpha_{1}\ket{1}_{\overrightarrow{r}}$$ and the discrete modelling  of instants $t_{1}, t_{2},...\ t_{N}$ will be $$\alpha_{0}\odot_{i=0}^{N}\ket{0}^{i}_{\overrightarrow{r}}+\alpha_{1}\odot_{i=0}^{N}\ket{1}^{i}_{\overrightarrow{r}}.$$By analog, in the spin $\sigma_{\overrightarrow{r}}$ view, we will immediately see that under the above setting, once we measure $\sigma_{\overrightarrow{r}}$ at some earlier instant, we will get the same value absolutely about the same spin for later measurements. And if we measure some part of the whole system, we will not get any information about the original state from any measurement of other subsystems.

 \

Note that in the above reasoning, the spin $\sigma_{\overrightarrow{r}}$ can be any spin, so we get  that in the above setting , for any spin, if we measure it at an earlier time $t_{k}$ and then measure it again at some later instant  $t_{l}, k\leq l$, then we will definitely get the same value for these two measurements. And this modelling will not tell too much about the original state. And this modelling can be easily generalized to general Hilbert spaces and unitray operators, once the orthonormal basis of each Hilbert space is given.

\

Until now, we give a definite answer to the question in \cite{yak}, using just the familar tensor product structure of Hilbert spaces. A vector in $\odot_{i=0}^{N}H_{i}$ can represent the discretization of a system's evolution.

\

Note that in the above modelling, $$\alpha_{0}\displaystyle\odot_{i=0}^{N}\ket{0}^{i}_{\overrightarrow{r}}+\alpha_{1}\odot_{i=0}^{N}\ket{1}^{i}_{\overrightarrow{r}},$$ measurements of this state and the corresponding collapse are just analogy. In \cite{frank4}, using monitor systems, we can transfer the temporal structure into the common spatial structure of those monitor systems. So actually, measurements and collapse occur on monitor systems. For details, \cite{frank4} can offer the beautiful correspondence.

On the other  hand, in the famous Leggett-Garg inequality \cite{LG}, like the Bell inequality \cite{bell, CHSH}, there are two fundamental principles behind it, one is macroscopic realism and the other is noninvasive measurability \cite{em}. The first says that a measurement should reveal a well-defined pre-existing value of a system, the second says that measurement will not disturb the studied system . Quantum violates both, superposion for the fist and collapse for the second. So, quantum theory violates the Leggett-Garg inequalities.

\

Before the entangled history formalism, there are already many efforts to unify the Bell inequality and the Leggett-Garg inequality \cite{mpj, ssr}. But unlike the Bell inequality  to the spatial nonlocality, for the Leggett-Garg inequality, the temporal nonlocality suffers great argument due to the strong effect of operators. And it is also important to see what consequences coherence results in a system's evolution way.  The entangled history formalism nicely shows this. And using the entangled history method, it can be easily seen that temporal nonlocality refers to the superposition of systems' evolution paths. Coherence of spatial states produces the spatial nonlocality and coherence of evolution path produces the temporal nonlocality.

\

In the classic physics, if we know the trajectory of a particle, then we can know its position and momentum at any instant. This means that from the  trajectory of a particle, we can know everything about. The reason is that we can measure  position and momentum of a particle exactly at the same time and  all observables are just functions of position and momentum. So in the classic case, position and momentum are natural coordinates to signify evolutions of a particle. However, in quantum physics, things are changed. We are not allowed to measurement all observables exactly at the same time. So in quantum physics, when we signify the evolution path of a system, we have to emphasize which  orthonormal basis we are using. And once orthonormal basis are fixed, we can only get information of those observables compatible with our orthonormal basis.

\

In \cite{yak}, they try to maintain the role of states as our common cases, from which we can know everything  about the process. However, as sacrifice,  they  hidden the impact of coherence and evolution's linearity on quantum evolution pictures, which are very important and distinguishable for quantum mechanics. In the entangled history method, physical meaning of quantum evolution is much clearer and more acceptable.

\subsection{the Energy-Time uncertainty relation}

 Uncertainty relationship of observables is one of most appealing traits of quantum mechanics. And this relationship has gained much discussion, from the physical view, from the informational entropy view and other different ways. But there is a bizarre uncertainty relationship, the Energy-Time uncertainty relation. Because in the usual form, time is not an observable (hermitian operator) in quantum mechanics, so it's very hard to describe this uncertainty relation in quantum information area,  which focuses on the finite-dimensional case. Recently, \cite{pvs} does a great work in this attempt. The core is  how to describe time uncertainty. In \cite{pvs}, they use skills to transform time uncertainty into discrimination of quantum states, which  inspires us very much. Below, we will describe the energy-time uncertainty relation in the entangled history formalism, from two extreme cases. The result is encouraging, and deserved to be studied seriously.

 \subsubsection{Energy is  fixed}

 Firstly, assume that our hermitian operator is non-degenerated. Then if the initial state of the system, $\ket{\phi}$ has a fixed energy, an eigenvector of our hermitian $H$. Now suppose that evolution time under this hermitian operator $H$ is $T$. Choose $N$ different instants $$t_{1},...\ t_{N}$$ from $(0, T]$ with the initial time $t_{0}=0$. Then the history space for our system is $\odot_{i=0}^{N}H_{i}$. If we use eigenvectors of the hermitian $H$
as the orthonormal basis of every $H_{i}$, then evolution of $\ket{\phi}$ is described as $$\ket{\phi}\odot \ket{\phi}\odot ...\odot \ket{\phi}.$$ It is clear that for states like $$\ket{\phi}\odot \ket{\phi}\odot ...\odot \ket{\phi},$$ it is impossible to find a measurement to tell which subsystem it is from measurement results. That means that in this  case, we can't make any inference about time, just can do random guessing. So in this case. the time uncertainty is biggest.

So, using the entangled history formalism, we show  that if a state has a fixed energy about the evolution hermitian ,which means that the energy uncertainty is  zero, then its time uncertainty must be biggest.

\subsubsection{Time is fixed}

As done in \cite{pvs},  time uncertainty is transformed into discrimination of states. In the entangled history formalism, we think that the sentence, time is fixed, means that we can find a measurement and from its outcomes, we can tell which subsystem it is perfectly. Let us use qubit system to show our idea. And to make it easier, the evolution only involves two instants $t_{0}$ and $t_{1}$. So in this setting, the history space is $C^{2}\odot C^{2}$. Suppose that our initial state is $\ket{0}$ and the unitary evolution between $t_{0}$ and $t_{1}$ is $X$. Then at the instant $t_{1}$, our system will be in the state $\ket{1}$, and the history state in this case is $$\ket{1}\odot\ket{0}.$$ Of course, in this  case, from measurement outcomes of $\sigma_{z}$, we can definitely tell which subsystem the measured qubit belongs to. This  means  that in this setting, there is  no time uncertainty. However, note that the uncertainty about the energy, the uncertainty of probability distribution measured by $\{\ket{+},\ket{-}\}$, is biggest, whether the state is $\ket{0}$ or $\ket{1}$. That is, when time is fixed, the uncertainty of energy is biggest.

\

The above reasoning is quite immature. But from the extreme cases, it seems that through the entangled history formalism, the energy-time uncertainty relation can get a very nice representation, which encourage us very much.

\subsection{Generalization to Complex Cases}

For methods in \cite{yak}, it works very well for pure states and unitary evolutions. But once states are mixed states or evolutions are general quantum operations, this method will be become very clumsy. Based on \cite{frank4}, we a  better way to deal with the general case.
\

Let's just talk about the two-instant setting. Now, suppose we have a system initialized in a state $\rho$ at the instant $t_{0}$, and it goes through the evolution $\Lambda$ until the instant $t_{1}$. In the above setting, $\rho$ is a general density matrix and $\Lambda$ is a general quantum operation. Methods in \cite{yak} works very well for vector states and general unitary evolution. But to deal with the above problem, it has to expand the original Hilbert space to purify $\rho$ and $\Lambda$, and then continues.  And what's worse, the partial trace operation is very difficult to define in the two-vector formalism. However, in the entangled history method, the construction is quite direct. Just as the simple, firstly, we fix the orthonormal basis $\{\ket{\alpha_{i}}\}_{i}$ and $\{\ket{\beta_{j}}\}_{j}$ for Hilbert spaces $H_{0}$ and $H_{1}$ corresponding to instants $t_{0}$ and $t_{1}$.  Then use $$E_{ij}=\ket{\alpha_{i}}\bra{\alpha_{j}},\ F_{kl}=\ket{\beta_{k}}\bra{\beta_{l}}.$$ Thus we will have the expression$$\rho=\sum_{i,j}\rho_{ij}E_{ij}.$$

In the simple case, when we fix orthonormal basis, we can give a matrix representation of the unitary operator. Similarly, with $\{E_{ij}\}_{ij}$ and $\{F_{ij}\}$, we can also give a matrix representation of a general quantum operation, just its Choi matrix, whose element is given by $$\Lambda_{kl,ij}=\bra{F_{kl}}\Lambda\ket{E_{ij}}=tr(F_{kl}^{\dagger}\Lambda(E_{ij})).$$

With these, the two-instant case for the state $\rho$ can be given directly $$\sum_{ij,kl}\Lambda_{kl,ij}\rho_{ij}F_{kl}\odot E_{ij}.$$ This generalization picture from vectors to matrices, from unitary matrices to Choi matrices is quite natural, with no need to expand the Hilbert space.  Using techniques in the entangled history formalism, we can  go directly into complex situations.  This is another trait of the entangled history method, compared with methods in \cite{yak}.

\section{Conclusion}

Superposition and linearity are the most important and fundamental features of quantum mechanics. From these two traits, we get a set of no-go theorems , such as the No-cloning theorem \cite{WZ}, the No-deleting theorem \cite{pati}. These  make quantum mechanics  depart quite far from the classic world. But usually, we refer superposition to quantum states at some fixed instant and linearity to the evolution  way  of quantum systems. In the entangled history formalism, they are combined together and through this combination we see the superposition of evolution paths of quantum systems and can get a better understanding of temporal  correlations. This shows  that the entangled history formalism  should play an important role in quantum theory. What's more, in \cite{NCH}, they show that there is a  isometric map between the two-vector formalism and the entangled history  formalism.  So if we see them as two equivalent formalism, then it's quite natural to ask the entangled history formalism to give an answer to the same question \cite{yak} dealt by the two-vector formalism.  In this paper, we give this answer , which is closer to their original intention. And furthermore, we try to give explanations of the Energy-Time uncertainty relationship from the entangled history formalism. At last compared with their work \cite{yak}, we find this solution is much easier to be generalised to complex cases.

\begin{acknowledgments}

\end{acknowledgments}


\begin{thebibliography}{10}

\bibitem{yak} Y. Aharonov, S. Popescu and J. Tollaken {\em Each Instant of Time a New Universe},
in Quantum Theory: A Two-Time Success Story, Yakir Aharonov Festschrift, edited by D. C. Struppa and J. M. Tollaksen (Springer, New York, 2013), pp. 21-36.
\bibitem{yak2} Y. Aharonov, S. Popescu, J. Tollaksen, L. Vaidman, {\em Multiple-time states and multiple-time measurements in quantum mechanics}, Phys. Rev. A {\bf 79}, 052110 (2009)
\bibitem{yak3} Y. Aharonov, L. Vaidman, {\em The two-state vector formalism of quantum mechanics in Time
in Quantum Mechanics}, in Time in Quantum Mechanics, edited  by J. G. Muga, R. Sala Mayato and I. L. Egusquiza (Springer 2002), pp. 369-412.
\bibitem{frank1} J. Cotler, F. Wilczek {\em Entangled Histories}, Phys. Scripta,
{\bf T168}, 014004 (2016).
\bibitem{frank2} J. Cotler, F. Wilczek, {\em Bell testss for Histories}, preprinted quant-ph/1503.06458.
\bibitem{frank3} J. Cotler, Lu-Ming Duan, Pan-Yu Hou, F. Wilczek, Da Xu, Zhang-Qi Yin, Chong Zu, {\em Experimental test of Entangled Histories} , Annals of Physics {\bf 387}, 334-347 (2017)
\bibitem{LG} A. J. Leggett, A. Garg, ¡°Quantum mechanics versus macroscopic
realism: Is the flux there when nobody looks?¡± Phys.
Rev. Lett. {\bf 54}, 857 (1985).
\bibitem{frank4} J. Cotler and F. Wilczek, {\em Temporal Observables and Entangled Histories}, preprinted quant-ph/1702.05838.
\bibitem{bell}J. S. Bell and A. Aspect, ¡°Speakable and Unspeakable in Quantum
Mechanics,¡± (Cambridge University Press, 2004).
\bibitem{CHSH} J. F. Clauser, M. A. Horne, A. Shimony, and R. A. Holt, ¡°Proposed
Experiment to Test Local Hidden-Variable Theories,¡±
Phys. Rev. Lett. {\bf 23}, 880 (1969).
\bibitem{em} C. Emary, N. Lambert, F. Nori, {\em Leggett-Garg Inequalities}, Rep. Prog. Phys. {\bf 77}, 016001 (2014)
\bibitem{ssr}S. Das, S. Aravinda, R. Srikanth, and D. Home,{\em Unification of Bell, Leggett-Garg and Kochen-Specker inequalities: Hybrid
spatio-temporal inequalities}, Europhys. Lett. {\bf 104}, 60006 (2013).
\bibitem{mpj} M. Markiewicz, P. Kurzy¡änski, J. Thompson, S.-Y. Lee, A. Soeda,
T. Paterek, and D. Kaszlikowski, {\em Unified approach to contextuality,
nonlocality, and temporal correlations}, Phys. Rev. A {\bf89},
042109 (2014).
\bibitem{WZ}  W.K. Wootters, W.H. Zurek,{\em A Single Quantum Cannot be Cloned}, Nature {\bf299} pp. 802-803  (1982),
\bibitem{pati} A. K. Pati, S. L. Braunstein , {\em Impossibility of deleting an unknown quantum state}, Nature {\bf404} pp.164¨C165 (2000)
\bibitem{NCH} M. Nowakowski, E. Cohen, and P. Horodecki, {\em Entangled histories versus the two-state-vector formalism: Towards a better understanding of quantum temporal correlations}, Phys. Rev. A {\bf 98}, 032312 (2018).
\bibitem{pvs} P. J. Coles, V. Katariya, S. Lloyd, I. Marvian, and M. M. Wilde, {\em Entropic Energy-Time Uncertainty Relation}, Phys. Rev. Lett. {\bf 122}, 100401 (2019)
\end{thebibliography}
\end{document}